# Dark Ages Radio Explorer Mission: Probing the Cosmic Dawn

Dayton L Jones
Jet Propulsion Laboratory
4800 Oak Grove Drive
Pasadena, CA 91109
818-354-7774
dayton.jones@jpl.nasa.gov

T. Joseph W. Lazio
Jet Propulsion Laboratory
4800 Oak Grove Drive
Pasadena, CA 91109
818-354-4198
joseph.lazio@jpl.nasa.gov

Jack O. Burns
CASA, 593 UCB
University of Colorado
Boulder, CO 80309
303-735-0963
jack.burns@colorado.edu

*Abstract*—The period between the creation of the cosmic microwave background at a redshift of ~1000 and the formation of the first stars and black holes that re-ionize the intergalactic medium at redshifts of 10-20 is currently unobservable. The baryonic component of the universe during this period is almost entirely neutral hydrogen, which falls into local regions of higher dark matter density. This seeds the formation of large-scale structures including the cosmic web that we see today in the filamentary distribution of galaxies and clusters of galaxies. The only detectable signal from these dark ages is the 21-cm spectral line of hydrogen, redshifted down to frequencies of approximately 10-100 MHz. Space-based observations of this signal will allow us to determine the formation epoch and physics of the first sources of ionizing radiation, and potentially detect evidence for the decay of dark matter particles.

JPL is developing deployable low frequency antenna and receiver prototypes and calibration techniques to enable both all-sky spectral measurements of neutral hydrogen at and ultimately to map the spatial distribution of the signal as a function of redshift. Such observations must be done from space because of Earth's ionosphere and ubiquitous radio interference. Both lunar orbiting and lunar surface based instruments are under development. A specific application of these technologies is the Dark Ages Radio Explorer (DARE) mission. This small Explorer class mission is designed to measure the sky-averaged hydrogen signal from the shielded region above the far side of the Moon. Observations of neutral hydrogen from the dark ages and the cosmic dawn (when light from the first generation of stars appeared) will provide unique and critical information on the early evolution of the universe. These data will complement ground-based radio observations of the final stages of intergalactic re-ionization at higher frequencies. DARE will also serve as a scientific precursor for space-based interferometry missions to image the distirbution of hydrogen during the cosmic dark ages.

TABLE OF CONTENTS

## 1. INTRODUCTION

The comic Dark Ages existed for a significant fraction of the first billion years since the Big Bang. This is a unique period between recombination (when the intergalactic medium cooled sufficiently due to cosmic expansion to become neutral) to the epoch of reionization (when the first generation of stars and black holes re-ionized the intergalactic medium). The epoch of recombination occurred near a redshift (z) of 1000, about $10^5$ years after the Big Bang, and produced the ubiquitous cosmic microwave background radiation. The epoch of reionization occurred between redshifts of 20 and 6. During the Dark Ages the universe was composed almost entirely of dark matter and neutral hydrogen. Eventually as hydrogen fell into local regions of higher dark matter density it became self-gravitating and collapsed to form the first stars. This process is illustrated in Figure 1.

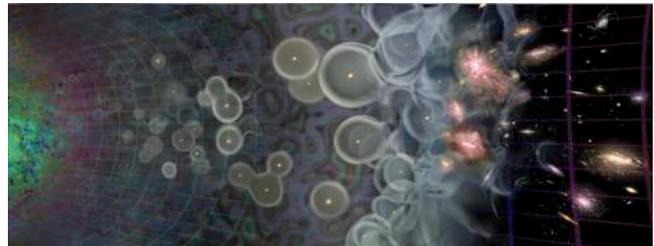

Figure 1 – Illustration of cosmic structure formation from the Big Bang (left) to the present epoch (right) [1]. The middle region of the figure shows ionized regions surrounding the first compact objects, and the eventual overlap and merging of these regions to produce today's highly-ionized intergalactic medium. Figure produced by Jean-Francois Podevin, used by permission.

During the Dark Ages, the redshifted 21-cm spectral line of neutral hydrogen is the only signal available for observing the formation of the large-scale structures that evolved into the cosmic web of filaments and voids we see in the current distribution of galaxies and clusters of galaxies. Because of the large redshifts, the 21-cm line from the Dark Ages will be observed at frequencies below ~100 MHz instead of its laboratory frequency of 1420 MHz (see Figure 2). These low frequencies are difficult or impossible to observed from the ground due to radio interference and the effects of refraction and absorption by Earth's ionosphere.

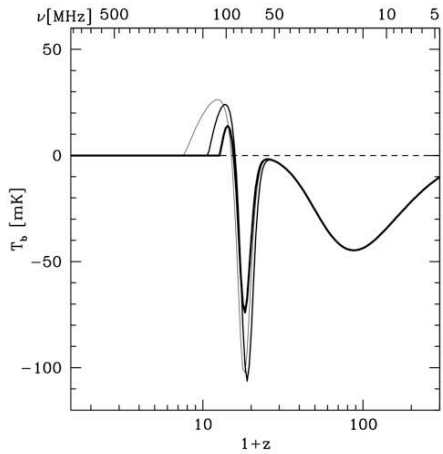

**Figure2 – Theoretical prediction of the sky-averaged neutral hydrogen spectral signal as a function of redshift (bottom axis) and observing frequency (top axis). The different curves represent different assumptions about heating and reionization of the intergalactic medium [2].**

The predicted amplitude of the neutral hydrogen emission and absorption shown in Figure 2 is typically a few tens of mK. This must be comparted to the known amplitude of foreground signals, primarily the synchrotron emission from cosmic ray electrons in our Galaxy. The Galactic foreground has a brightness temperature of thousands of degrees at these frequencies. Thus, we are faced with a signal-to-noise ratio of approximately $10^{-6}$.

The hydrogen signal will be, to a large degree, spatially uniform across the sky while the galactic synchrotron emission is strongly concentrated towards the galactic plane. In addition, the hygrogen signal is spectrally complex while synchrotron emission is spectrally smooth. Consequently it is possible to detect the faint hydrogen signal buried in the strong foreground signal. However, this requires excellent instrumental stability and careful attention to calibration and systematic errors.

At low radio frequencies the Earth's ionosphere becomes opaque. Even above that point the ionosphere introduces severe absorption, emission, and refractive effects to incoming astronomical radio signals. In addition, the ionosphere can propagate strong time-varying interfering signals from distant locations to anywhere on Earth's surface.

To minimize systematic errors in observations of highly redshifted hydrogen we need to be above Earth's ionosphere and sheilded from both natural and human-generated interference. To accomplish this we plan to observe from a spacecraft in low lunar orbit, using periods when the Moon shields the spacecraft from both Earth and solar noise and provides a thermally stable nighttime environment. The far side lunar surface and the space above it represent the best location anywhere in the inner solar system for these observations. An additional advantage is that the Moon is not so distant that data downlink to Earth becomes difficult.

## 2. SCIENCE INSTRUMENT FOR THE DARK AGES RADIO EXPLORER MISSION

The basic components of the Dark Ages Radio Explorer (DARE [3]) science instrument being developed at JPL are described in the following subsections.

*Antenna*

The DARE antenna is a bi-conical dipole. This geometry is mechanically and electrically simple, provides a wider fractional bandwidth than a skinny dipole, and has a nearly uniform beam at frequencies below resonance. A test model of this antenna geometry and a prototype deployable version are shown in Figure 3. The RF performance of both were measured in an electromagnetic shielded room at JPL.

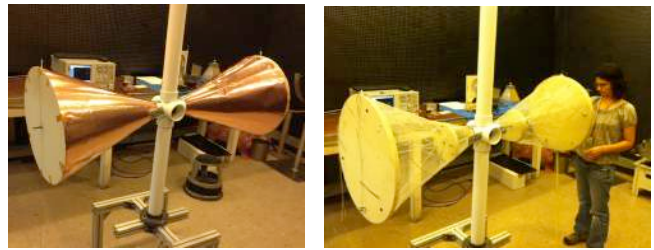

**Figure 3 – Fixed reference biconical dipole antenna (left) and a prototype deployable antenna for DARE (right).**

The DARE spacecraft will carry a pair of orthogonal dipole antennas, as shows in Figure 4. This provides redundancy, and potentially a square-root-of-two increase in sensitivity.

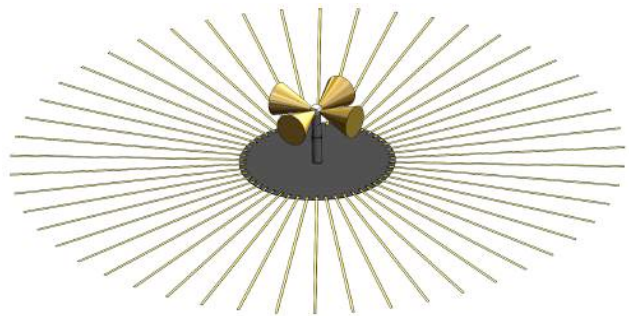

**Figure 4 – DARE antenna system after deployment, including radial ground plane. The spacecraft body is located under the solid inner portion of the ground plane. (Figure produced by J. Sauder at JPL.)**

*Receiver*

Achieving the thermal noise sensitivty needed to detect the redshifted hydrogen signal is not difficult. The challenge is to reduce systematic errors to a level where they do not prevent a reliable measurement of the spectral shape of the hydrogen signal. One aspect of this is to reduce the effect of gain variations in the receiver (1/f noise). We employ the correlation receiver architecture shown in Figure 5 to minimize sensitivty to gain changes [4, 5].

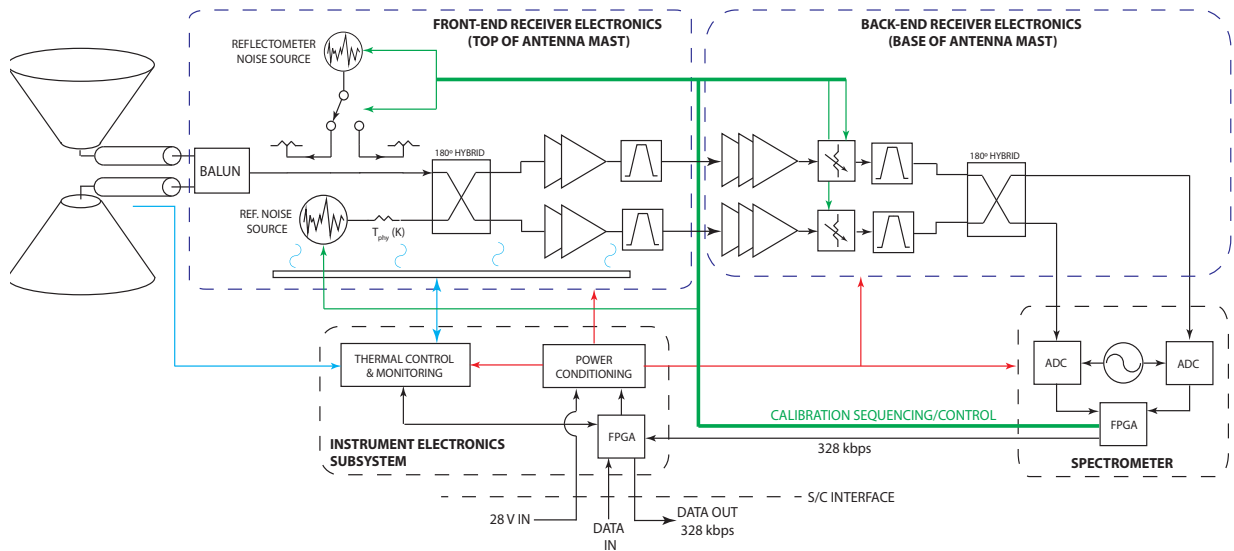

**Figure 5 – DARE pesudo-correlation receiver architecture (figure provided by D. Russell at JPL).**

Another source of error is the frequency-dependent antenna reflection coefficient. Because DARE will observe over a three-to-one frequency range, the antenna impedance will vary dramatically and at some frequencies will present a poor match to the receiver input. In this case even small change in the antenna match can introduce spectral features that could mask (or be mistaken for) the hydrogen signal. To account for this the physical temperature of the antenna will be monitored so changes in geometry can be calculated. In addition, the receiver includes a built-in reflectometer to measure the antenna reflection coefficicnt periodically.

*Spectrometer*

The analog output from the receiver is digitized and goes to an FPGA-based digital, polyphase filter bank spectrometer that produces spectra with 4096 frequency channels every few seconds. In addition to total power, the spectrometer also calculates the fouth moment (kurtosis [6]) to detect departures from gaussian noise statistics. A gaussian distribution has a kurtosis of 3, while most interference or instrumental artifacts will produce kurtosis values greater than 3. A breadboard version of the spectrometer during a long-integration bench test of stability and spectral dynamic range is shown in the right panel of Figure 6.

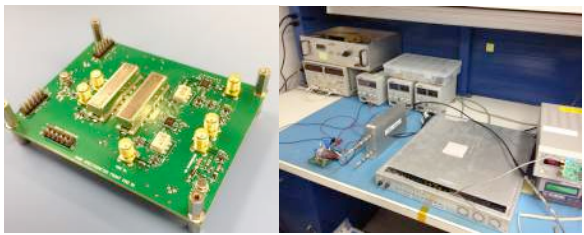

**Figure 6 – Spectrometer calibration board (left) and long-integration testing of the spectrometer (right).**

*Calibration*

The end-to-end system calibration is based on a broad-band noise source that is switched at a 1 Hz rate. The stability of this noise source is critical, and the output will be calibrated against the sky in the direction of the galactic center at regular intervals. There is also higher level noise source as part of the reflectometer that can be used periodically to measure the absolute gain and system temperature of the receiver. The front-end receiver electronics are temperature controlled in a separate thermal enclosure, and the back-end electronics are located within a temperature controlled area of the host spacecraft. However, it is not possible to actively control the physical temperature of the antenna. Variations in the antenna impedance as a function of physical temperature are simulated using the measured temperatures and antenna modeling software. The antenna simulations are based on pre-flight laboratory measurements of the antenna characteristics across the frequency range at a number of different physical temperatures. During flight the antenna beam pattern will be verified using the unresolved radio source Casseopeia A.

Calibrated spectra from eight independent sky directions measured by the DARE science instrument are used in a Markov Chain Monte Carlo (MCMC) parameter fitting program to simultaneously solve for the spectrum of the all-sky hydrogan signal, the galactic foreground signal, and instrument parameters [7]. The MCMC code provides estimates of the frequencies and amplitudes of the spectral inflection points seen in Figure 2.

*Testing*

The antenna needs to be physically larger than the expected launch volume available in order to provide acceptable performance at the low end of the DARE frequency range. Consequently the antenna must deploy after launch, and rigorous verification of deployment mechanism reliability including thermal, vacuum, and vibration tests will be carried out at JPL.

The spectrometer dynamic range has been measured with a test board that produces broad-band noise with a spectral index matched to the galactic synchrotron foreground radiation and a weak spectral freature added to simulate the expected hydrogen signal. The spectrometer test board is shown in the left image of Figure 6, and its architecture is shown in Figure 7.

A full end-to-end test of instrument performance from the antenna through data out to the spacecraft telemetry system will be carried out prior to integration with the spacecraft. As mentioned before, systematic errors in the spectral data are out biggest concern, and establishing the level of these errors will be the primary focus of the end-to-end tests.

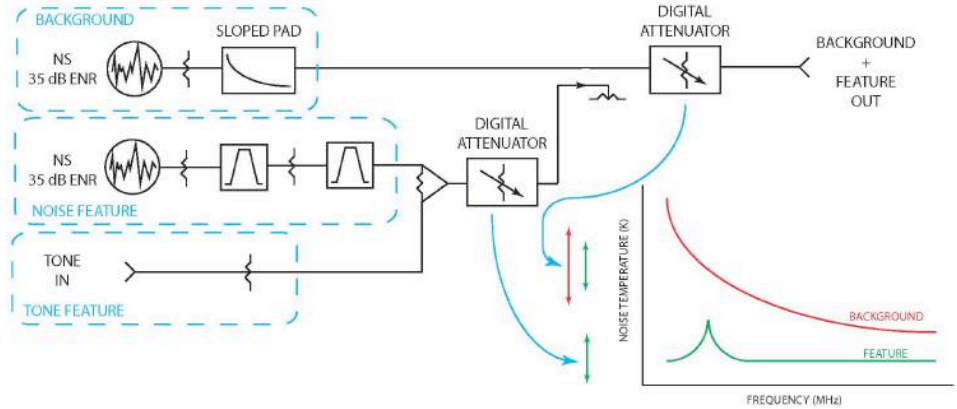

**Figure 7 – Spectrometer dynamic range test signal.**

Results from a long-integration test of the spectrometer are shown in Figure 8. These tests are continuing; the DARE mission will observe the sky from the double-shielded (both Earth and Sun eclipse) portion of its lunar orbit for at least 1000 hours, excluding calibration time.

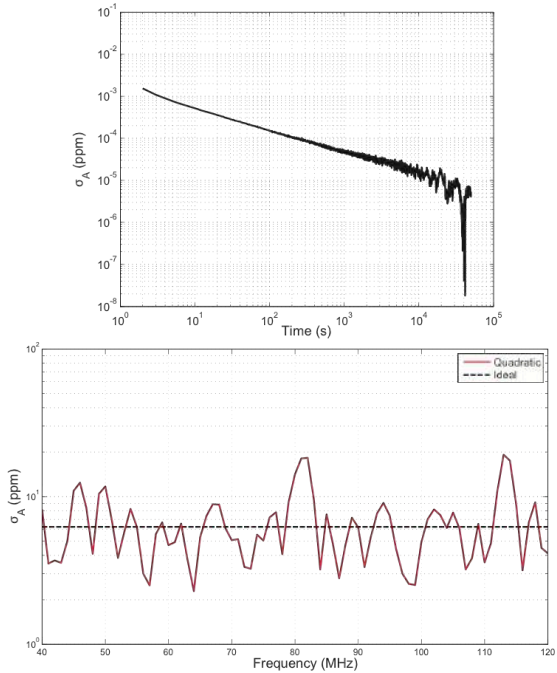

**Figure 8 – The measured Allan variance vs time (top), and the Allan variance in 64-channel averages across the full instrument frequency range (bottom) produced by the DARE spectrometer after 72 hours of integration.**

## 3. ASSOCIATED TECHNOLOGY DEVELOPMENT

The DARE mission will measure the all-sky spectral signal from hydrogen between redshifts of approximately 35 and 12. This will tell us the epochs at which sources of ionizing radiation formed, and will provide constraints on the physical conditions. A longer-term and more challenging goal is to measure the angular distribution of the signal as a function of redshift. This can be done through observation of the angular correlation function or some other statistical measure of angular structure. These observations will require an interferometer.

The ultimate goal of producing a detailed three-dimensional view of large-scale structure evolution during the dark ages and cosmic dawn from high dynamic range images of the hydrogan signal at many redshifts will provide the greatest information about these epochs, but will require a very large and sensitive space-based interferometer.

A space-based radio interferometer, especially if located at the optimal lunar far-side surface location (shielded from terrestrial interference all the time and from solar radio bursts half the time), will require a digital cross-correlator to reduce the bandwidth of data downlinked to Earth. The correlator will have to operate in a highly power-constrained environment during lunar night. JPL has invested in the design of an application specific integrated circuit (ASIC) chip for digital cross-correlation with extremely low power consumption [8, 9]. This chip may also be relevant for future large ground-based radio interferometers, including low frequeny arrays to study the epoch of reionization and the Square Kilometre Array. The chip can correlate all signals (and both polarizations) from radio arrays having between 32 and 4096 antennas. The number of antennas is traded against the total bandwidth that can be processed on a single chip.

Other areas of current development at JPL include low radio frequency, low mass antennas for eventual deployment on the lunar surface [10, 11]. The future lunar-based arrays needing hundreds of antennas, each antenna must have a mass of grams, not kilograms. One way to achieve this is to use thin, metal-coated polyimide film for the antenna elements. Both rover-deployed and inflation-deployed long dipole antennas of this sort have been demonstrated successfully in the JPL Mars Yard (see Figure 9). Further RF performance testing of the antennas is underway at nearby desert sites with less local interference and dryer ground conditions than are available at JPL.

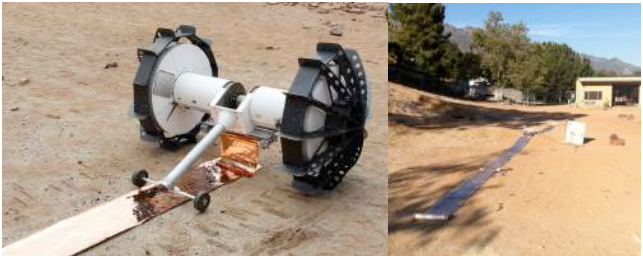

**Figure 9 – Polyimide film antenna deployed by an Axel rover (left), and a similar antenna after an inflation-driven deployment test (right). These long dipole antennas are intended for observations below ~10 MHz.**

The goal of this work is to prepare for low radio frequency observations from space in the post-DARE era.

## 4. FUTURE WORK

*Near-term*

Besides the DARE mission, there are a number of ground-based experiments underway to measure the lower redshift hydrogen signal from the epoch of renionization. These include single-antenna instruments (e.g., EDGES [12]) and interferometers with tens to hundreds of antennas (e.g., LOFAR [13], MWA [14], and PAPER [15]). A proposed new interferometer, the Hydrogen Epoch of Reionization Array (HERA), is under development. The JPL low power correlator ASIC could reduce the operating cost of this and other future arrays.

Technology development aimed at space-based low radio frequency interferometers following the DARE mission will contine. As with the cosmic microwave background (CMB), ground-based observations will inevitably be followed by space observations to remove the limitations imposed by the ionosphere and terrestrial interference. Among the current concepts for space-based interferometry missions are the Orbiting Low Frequency Antennas for Radio Astronomy (OLFAR [16]), the Distributed Aperture Array for Radio Astronomy in Space (DARIS, [17]), the Space-based Ultra-long wavelength Radio Observatory (SURO [18]), PrepAration for Radio Astronomy in Space (PARIS [19]), and a number of other concepts under study in India, China, and the USA.

There are other science goals that require observations at even lower frequencies, such as imaging solar radio bursts [20], searches for magnetospheric radio emission from exoplanets [21], and studies of the (tenuous and transitory) lunar ionosphere [22]. Much of the technology needed for cosmological hydrogen observations can be applied at lower frequencies; the only fundamental difference is the physical size of the antennas needed.

*Long-term*

The huge information content of high-redshift hydrogen, which unlike the CMB is available over a wide range of redshifts, can be fully exploited only with a very large space interferometer located in a radio quiet location. The lunar far side is the best site in the inner solar system for such an instrument. Consequently, continuing development work to make a large lunar interferometer feasible in terms of mass, power consumption, downlink data rates, and cost will be needed. JPL intends to continue development of appropriate technlgies in order to be ready for future mission proposal opportunities.

## 5. CONCLUSIONS

The Dark Ages Radio Explorer will open a new window for cosmology by detecting the spectral line of hydrogen at redshifts greater than 15. Future space-based low frequency interferometers will built on the DARE measurements and will eventually give us a full view of the evolution of large scale structure formation leading to the cosmic web we see in the distribution of galaxies in the local universe.

## 6. ACKNOWLEDGEMENTS

This work is being carried out at the Jet Propulsion Laboratory, California Institute of Technology, under contract with the National Aeronautics and Space Administration. We are grateful for the hard work of the DARE proposal team at the University of Colorado, Ames Research Center, and JPL. We particularly thank D. Russell and C. Parashare for providing figure 5-8, and R. Reid for figure 9. Previous work leading to the DARE mission concept was partly supported by the Luanr University Node for Astronphysical Research (LUNAR [23]), a consortium funded by the NASA Lunar Science Institute.

## BIOGRAPHY

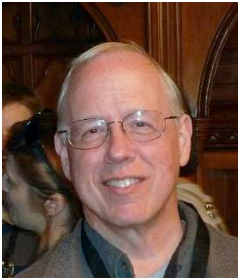

*Dayton Jones is a Principal Scientist at the Jet Propulsion Laboratory, California Institute of Technology. He has been involved in studies of space-based low frequency array mission concepts for the past two decades, most recently concepts for a large lunar-based radio array to observe neutral Hydrogen from early cosmic epochs. His research interests include high angular resolution imaging and high precision astrometry with very-long-baseline interferometry. He is an author on ~200 scientific publications. He has a BA in Physics from Carleton College, an MS in Scientific Instrumentation from the University of California, Santa Barbara, and MS and PhD degrees in Astronomy from Cornell University.*

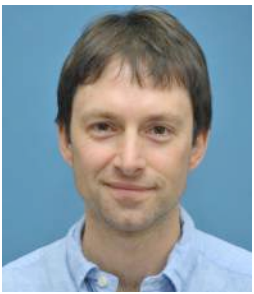

*Joseph Lazio is a Principal Scientist at the Jet Propulsion Laboratory, California Institute of Technology. His research interests include studies of the Dark Ages and Cosmic Dawn using the highly redshifted 21 cm line from neutral hydrogen and gravitational wave astronomy. He observes frequently with radio telescopes from around the world, and has been involved in the development of a number of concepts for new telescopes, including antennas and arrays on the Moon. He is the author or co-author of over 100 scientific publications. He has a B.S. in Physics with Highest Honors from the University of Iowa and M.S. and Ph.D. in Astronomy from Cornell University.*

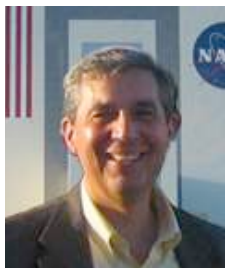

*Jack Burns is Professor of Astrophysics and Vice President Emeritus for Academic Affairs and Research at the University of Colorado, Boulder. He also directs the NASA Lunar Science Institute's Lunar University Network for Astrophysics Research (LUNAR), a national consortium of 10 universities and 3 NASA centers. He has led research studies on lunar-based observatories, especially low frequency telescopes, for 30 years. He is currently Principal Investigator of the Dark Ages Radio Explorer (DARE) mission concept project. He has over 380 publications. He received his B.S. in Astrophysics from the University of Massachusetts and his Ph.D. from Indiana University.*